# Influences of substrate thickness on neutron specular reflection

S. F. Masoudi

Department of physics, Tehran University, P.O. Box 1943-19395, Tehran, Iran.

**Abstract**
In specular reflection experiments the reflected beam from the end side of thick substrates is typically neglected. This is equivalent to assuming the substrates as semi-infinite matter. However, it is known that we should also consider the reflected beam from the end sides. Here we have investigated the effect of this consideration on the average reflectivity in a completely general case. It is shown that in some cases, especially in local minima of reflectivity vs. neutron wave number, this consideration can result in high enough differences such that it should be included in the interpretation of measured neutron specular reflectivity.

## 1. Introduction

When a neutron beam strikes a stratified sample, the neutrons interact with individual atomic nuclei as well as the magnetic induction generated by the atomic electrons [1,2]. The reflected ray in the specualr direction gives information on the refractive index profile (with respect to neutron beam) normal to the sample. This index is simply related to the Scattering Length Density (SLD) profile normal to the surface sample which, in turn, can be directly converted to the chemical and/or magnetic profiles of the sample [3]. So specular neutron reflectometry can provide important information about the composition of surfaces and interfaces [4-6]. The more accurate be interpretation of the measured data in specular neutron reflection, the more precise is the information we obtain.

In specular reflection experiments with neutrons, stratified thin films are mounted on top of a substrate with macroscopic thickness. So the measured data usually are analysed under the simplifying assumption that the substrate is a semi-infinite matter, i.e., it is infinitely thick. It is obvious that by this assumption the reflected ray from the end side of the substrate is neglected. This assumption is sometimes rational especially when the neutron current falls down considerably before reaches the end side of the substrate.

But for some materials as substrate with weak absorption, such as silicon [7], the reflection from the end side may be noticeable and have appreciable effects on the average reflectivity. Therefore to accurately interpret the results of specular reflection experiments, weakly absorbing substrates must be considered as a thick sample instead of treating them as semi-infinite matter. Effects of the reflected ray from the end side of substrate on average reflectivity have been considered firstly by Reiss and Lipperheide [8]. Here we will take into account these effects in a completely general case and show that it has some effects on the correct interpretation of the measured data in neutron specular reflection.

The layout of the paper is as follows. In section 2, we will calculate the average of the complex reflection coefficient and the average of the reflectivity in two cases; semi-infinite and thick substrates. It will be shown that the averages of the reflection coefficient in two cases are the same but the averages of the reflectivity are different. In section 3, we introduce some examples to examine this difference. By using the simplest example, i.e., a uniform thin film with a constant SLD on top of a substrate, it is shown that the difference is larger in the minima of the reflectivity. In the last example, we investigate the effects of reflection from end side on a quantity that used in determination of the phase of the reflection coefficient by using polarized incident neutron beam. The paper is ended by a conclusion.

## 2. Effect of substrate thickness

As a completely general sample for investigation by neutron specular reflection, consider a group of stratified microscopically thin films mounted on a macroscopically thick sample. The thin films are of the order of a micron with arbitrary profiles and the substrate is of the order of a millimeter or centimeter with a constant SLD. It is assumed that the neutrons entering the sample are not absorbed and also do not leave it unreflected through the edges.

This is easily seen that the expression for the reflection coefficient of the whole arrangement from the left can be expressed in terms of the reflection and transmission from left and right of the whole sample without substrate [9]

$$r_t = \frac{Er_s + r_L}{1 - r_R r_s}, \tag{1}$$

where

$$E = t_L t_R - r_L r_R, \tag{2}$$

and $r_{L(R)}$ and $t_{L(R)}$ are the reflection and transmission coefficients from left(right) by the stratified thin films mounted on a bulk substrate and $r_s$ is the reflection coefficient of the substrate. If we consider a constant SLD for the substrate, $\rho_s$, $r_s$ is equal to

$$r_s = r_F \frac{1 - \exp(2idq_s)}{1 - r_F^2 \exp(2idq_s)}, \tag{3}$$

where q is the incident wave number, $q_s = \sqrt{q^2 - 4\pi\rho_s}$ is the neutron wave number inside the substrate, d is the thickness of substrate, and

$$r_F = \frac{q - q_s}{q + q_s}, \tag{4}$$

is the Fresnel coefficient for the reflection at the left side of the substrate.

If we consider the substrate as a semi-infinite matter, $r_s = r_F$, then the reflection coefficient for the whole sample with semi-infinite substrate is

$$r_\infty = \frac{Er_F + r_L}{1 - r_R r_F}. \tag{5}$$

By using Eq. (3), Eq. (1) can be expressed as

$$r_t = r_\infty \frac{1 - \alpha \exp(2idq_s)}{1 - \beta \exp(2idq_s)}, \tag{6}$$

where

$$\alpha = \frac{Er_F + r_L r_F^2}{Er_F + r_L}, \tag{7}$$

$$\beta = \frac{r_F^2 - r_R r_F}{1 - r_R r_F}. \tag{8}$$

To calculate the averages over incident wave number, we write Eq. (6) in the form

$$r_t = r_\infty \left( \frac{\alpha}{\beta} + \frac{1 - \alpha/\beta}{1 - \beta \exp(2idq_s)} \right) \tag{9}$$

Assuming β to be slowly varying over the averaging interval $\Delta \gg \pi/d$, then we get $\langle 1/(1-x) \rangle = 1$ where $x = \beta \exp(2idq_s)$, hence

$$\langle r_t \rangle = r_\infty. \tag{10}$$

Eq. (10) shows that average of reflection coefficient for thick substrate is indeed equal to the reflection coefficient for semi-infinite substrate. Therefore the averaging has the effect of neglecting the reflection from the end side of substrate. But the complex reflection coefficient cannot be measured directly in experiments because as any scattering technique the phase information is lost. In fact what is measured is the average reflectivity. Now we calculate the effect of reflection from end side on this quantity.

By using Eq. (6) the reflectivity can be express as

$$R_t = |r_t|^2 = |r_\infty|^2 \left|\frac{\alpha}{\beta}\right|^2 \left\{ 1 + 2\text{Re}\left(\frac{\beta/\alpha - 1}{1 - \beta \exp(2idq_s)}\right) + \left|\frac{\beta}{\alpha} - 1\right|^2 \frac{1}{|1 - \beta \exp(2idq_s)|^2} \right\}. \tag{11}$$

So, we have

$$\langle R_t \rangle = |r_\infty|^2 \left|\frac{\alpha}{\beta}\right|^2 \left\{ 1 + 2\left\langle \operatorname{Re}\frac{\beta/\alpha - 1}{1-x} \right\rangle + \left|\frac{\beta}{\alpha} - 1\right|^2 \left\langle \frac{1}{|1-x|^2} \right\rangle \right\}, \tag{12}$$

The two last terms (averages) in the right hand side of Eq. (12) can be express as

$$\left\langle \frac{1}{|1-x|^2} \right\rangle = \left\langle \sum_{n,m=0}^{\infty} x^n x^{*m} \right\rangle = \sum_{n,m=0}^{\infty} \left\langle x^n x^{*m} \right\rangle = \sum_{n,m=0}^{\infty} |x|^{2n} \delta_{mn} = \sum_{n=0}^{\infty} |x|^{2n} = \frac{1}{1-|x|^2}, \tag{13}$$

$$2\left\langle \operatorname{Re}\frac{\beta/\alpha - 1}{1-x} \right\rangle = \left\langle \frac{\beta/\alpha - 1}{1-x} \right\rangle + \left\langle \left(\frac{\beta/\alpha - 1}{1-x}\right)^* \right\rangle = \left(\frac{\beta}{\alpha} - 1\right) + \left(\frac{\beta}{\alpha} - 1\right)^* = 2\operatorname{Re}\left(\frac{\beta}{\alpha} - 1\right). \tag{14}$$

Replacing these equations in Eq. (12) it is easy to find

$$\langle R_t \rangle = |r_\infty|^2 + |r_\infty|^2 \frac{|\alpha - \beta|^2}{1-|\beta|^2}. \tag{15}$$

This equation shows that in the direct problem of reflection, i.e., calculation of the reflectivity given SLD profile of the whole sample with thick substrate, the simulated reflectivity must be corrected by $\Delta R$,

$$\langle R_t \rangle = |r_\infty|^2 + \Delta R \tag{16}$$

where

$$\Delta R = |r_\infty|^2 \frac{|\alpha - \beta|^2}{1-|\beta|^2}. \tag{17}$$

By using Eqs. (7) and (8) the above relation gives the following expression

$$\Delta R = \frac{r_F^2 (1 - r_F^2) |t_R t_L|^2}{(1 + r_F^2 |r_R|^2 - 2r_F \operatorname{Re} r_R)(1 + r_F^2 - 2r_F \operatorname{Re} r_R)}. \tag{18}$$

## 3. Some examples

In this section we explicitly investigate the effect of reflection from the end side for two different examples. In the first example we consider a free substrate without any thin films. This is equivalent to say that we deal with a macroscopic sample. In the second one, at first we deal with a constant SLD thin film mounted on a substrate. Then we show the effects of the reflected beam on the average reflectivity. We go further and investigate a general stratified thin film in the same regard. As the last case, the same general arrangement of the second example supplemented with an external magnetic field is studied. There we show the effect of the reflected beam from the end side of the substrate on the polarization of the reflected beam from whole sample normal to the surface. Then we investigate this effect on an important parameter determined by polarization of the reflected beam and used in determination of the SLD profile of an unknown non-magnetic film by polarized incident beam.

### 3.1 Wafer
As first example, consider a thick sample that can be used as a substrate for thin films (wafer). Since there are no thin films, then $r_L = r_R = 0$ and $t_L = t_R = 1$, from which $r_\infty = r_F$ and

$$\Delta R = \frac{r_F^2 (1 - r_F^2)}{1 + r_F^2}. \tag{19}$$

In the other words, we have

$$\langle R_t \rangle = \frac{2 r_F^2}{1 + r_F^2}. \tag{20}$$

For the large values of q in which the Fresnel reflection is far less than unity, $r_F \ll 1$, we have

$$\langle R_t \rangle = 2 r_F^2. \tag{21}$$

This indicates that for large values of q the average reflectivity is twice as large as the Fresnel reflectivity [8]. In Fig. 1 as an example for a wafer, we consider a silicon wafer having positive constant SLD value of $2.08 \times 10^{-4}$ nm$^{-2}$ and show the difference between $<R_t>$ and $R_\infty = |r_\infty|^2 = r_F^2$. Fig. 2 illustrates the similar graph for a Ti wafer having negative constant SLD value of $-1.95 \times 10^{-4}$ nm$^{-2}$. The curves show that the reflected beam from the end side of silicon has no effect on critical edge.

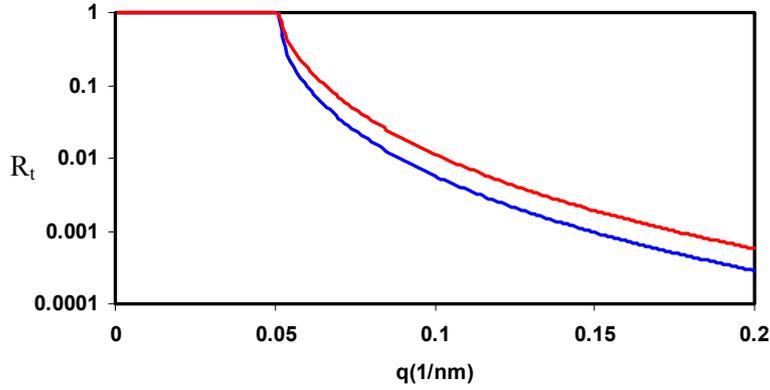

Fig. 1. The difference between the average reflectivity for thick silicon (red) and the reflectivity of semi-infinite silicon (blue).

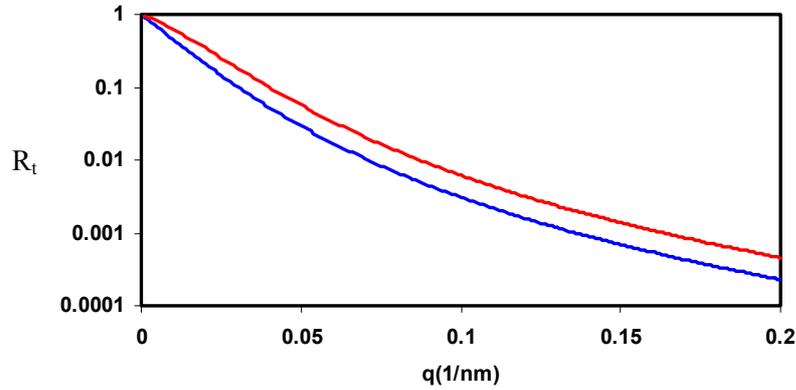

Fig. 1. The difference between the average reflectivity for thick Titanium and the reflectivity of semi-infinite one.

**3.2 General case**

A uniform layer with a constant SLD mounted on top of a substrate is the simplest form for general case. It is important because many experiment films can be regarded, to the first approximation, as a uniform film with an average SLD on top of a substrate. The reflectivity from such an average film is usually a dominant feature in the reflectivity from the actual samples. For example, the total reflection plateau for an actual sample is more or less the same as when the profile is replaced with a uniform film with the average SLD. Now some of general behaviors of reflectivities can be understandable. On the other hand, gas, liquid and amorphous solid have constant SLD. As an example for this case, consider a gold thin film with 50 nm thickness and $4.66\times10^{-4}$ nm$^{-2}$ SLD on top of a silicon substrate. Fig. 3 shows the difference between $<R_t>$ and $R_\infty$. As is seen in the local minima of the average reflectivity the difference can be appreciable, however this is not the case for other points. A simple explanation for this behavior can be the fact that by averaging some data we get smoother values. This ever may result in removing oscillating behavior of the reflectivity.

   As a further example, consider an experimental arrangement as shown in Fig. 4. It consists of a silicon substrate with 50-nm-thick gold layer ($\rho=4.66\times10^{-4}$ nm$^{-2}$) and 10-nm-thick cobalt layer with no magnetization ($\rho=2.23\times10^{-4}$ nm$^{-2}$). $R_\infty$, $<R_t>$ (Eq. (19)) and $<R_t>$ (as calculated in [4] ) are shown in this figure. The difference between Eq. (19) and the one calculated by Reiss and Lipperhiede [8] due to the fact that they did not use the complete expression for the reflection coefficient of the substrate, Eq. (3).

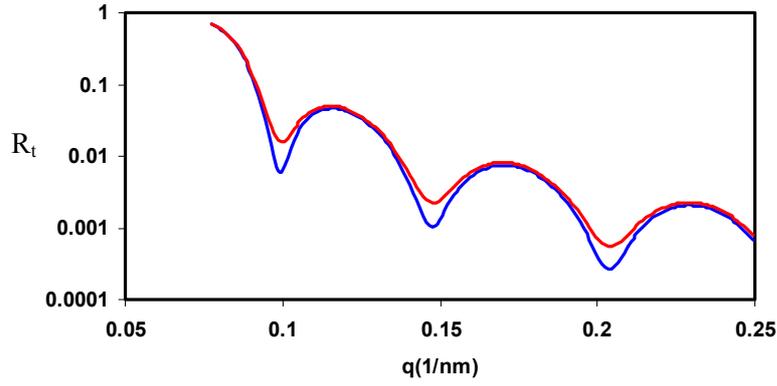

Fig. 3. The difference between $\langle R_t \rangle$ (red) and $R_\infty$ (blue) for 50 nm-thick gold layer mounted on top of a silicon substrate. The simulated data start at the critical neutron wave number of silicon.

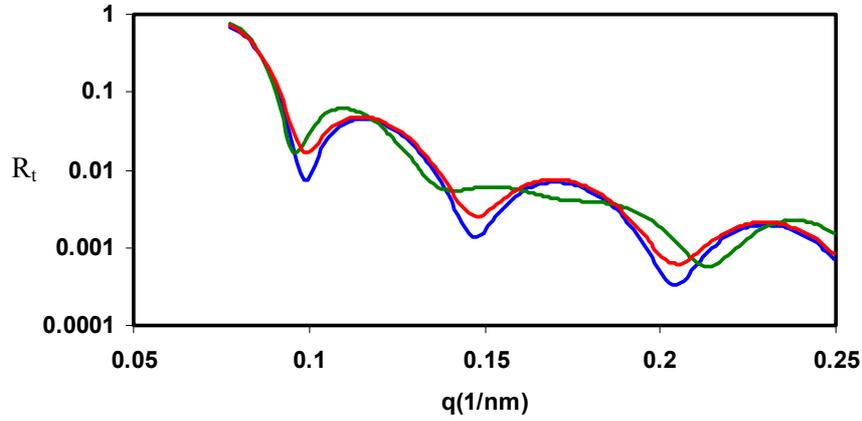

Fig. 4. $R_\infty$ (blue), $\langle R_t \rangle$ (Eq. (19)) (red) and $\langle R_t \rangle$ ([4]) (green) for the stratified thin films shown in the inset. The simulated data start at the critical q of silicon.

**3.3 Thin films under external magnetic field**

When a non-polarized incident neutron beam strikes a magnetic film under an external magnetic field, the reflected beam from whole sample is polarized. In this example we investigate the effect of reflection from the end side of substrate on the polarization of the reflected beam normal to the surface sample, $\sigma_\perp$.

For a magnetic film, two different magnetization due to an incident beams polarized in up and down direction lead to two different SLD's for the magnetic layer [10]. Two different SLD's cause two different reflection coefficient $r_t^+$ and $r_t^-$ respectively for up and down polarized incident beams. $\sigma_\perp$ is simply related to $r_t^+$ and $r_t^-$ as follow

$$\sigma_\perp = \frac{\left|r_t^+\right|^2 - \left|r_t^-\right|^2}{\left|r_t^+\right|^2 + \left|r_t^-\right|^2}, \tag{22}$$

The effect of the reflection from the end side of substrate on $\langle\sigma_\perp\rangle$ can be determined easily using Eq.(3). To illustrate the difference between $\langle\sigma_\perp\rangle$ and $\sigma_\perp$ for semi infinite substrate, consider the experimental arrangement shown in Fig. 4, but for two different magnetizations of the cobalt layer ($\rho^+ = 7.08 \times 10^{-4}$ nm$^{-2}$ and $\rho^- = -2.62 \times 10^{-4}$ nm$^{-2}$ respect to plus and minus magnetization), Fig 5 (This arrangement was presented in Refs. [11,12] to show the ability of retrieval of the phase information in neutron reflectometry, however, we mount non-magnetic layer on magnetic layer). The effect of reflection from the end side of substrate on $\sigma_\perp$ for this example shown in Fig. (6).

The polarized incident neutron beam and magnetic reference films can be used to determination of the SLD of an unknown film. The difference between the polarization of reflected neutron for semi-infinite substrate and thick substrate may have appreciable effect on determination of the SLD profile [13]. Here we investigate this effect on the absolute square of an important quantity in magnetic reference layer method of phase determination with incident polarized neutron, $|s|^2$ where [12,14],

$$s = \frac{r_t^+}{r_t^-}, \qquad (23)$$

This complex quantity can be determined experimentally by measuring the polarization of the reflected neutron in three directions. The absolute square of s depends only on $\sigma_\perp$ as follow,

$$|s|^2 = \frac{1+\sigma_\perp}{1-\sigma_\perp}, \qquad (24)$$

Fig. (7) depicts the difference between $\langle|s|^2\rangle$ and $|s|^2$ that determined by $\langle\sigma_\perp\rangle$ and $\sigma_\perp$ respectively.

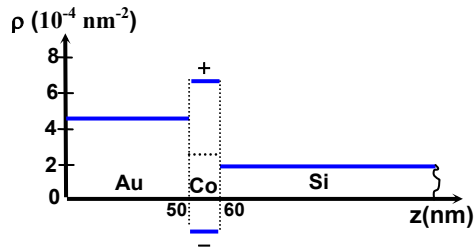

Fig. 5. The arrangement used in sec. 3.2, under an external magnetic field on the cobalt layer.

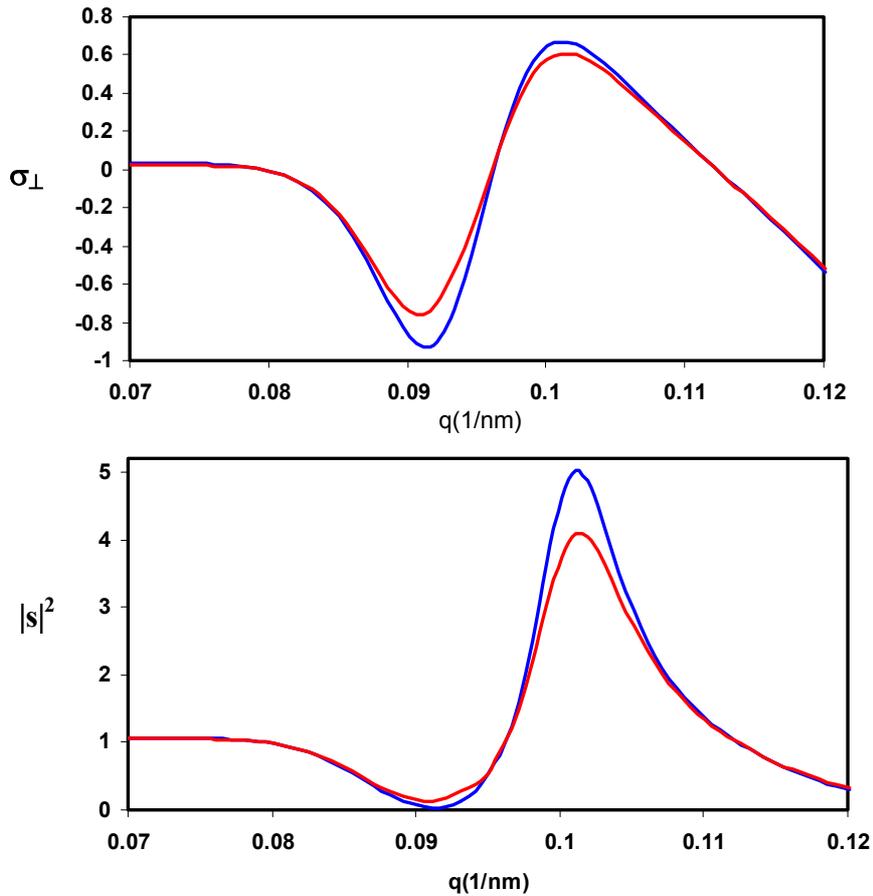

Fig. 6. Difference between $\langle\sigma_\perp\rangle$ (red) and $\sigma_\perp$ (blue) for the arrangement shown in Fig (5).
Fig. 7. The difference between $|s|^2$ calculated by considering the substrate as a semi-infinite matter (blue) and thick matter (red).

Fig. (7) shows that the changes is high enough to be taken into account seriously in determination of the phase of the complex reflection coefficient.

**4. Conclusions**

We have taken into account of the thickness of the substrates instead of treating them as semi-infinite substrates. This is shown to have appreciable effects on the average reflectivity, however, not on the reflection coefficient. The difference between the average reflectivity and the reflectivity is calculated for a completely general case. As an example, we have explicitly shown the effect of the reflection from the end side of the substrate on the polarization of the reflected beam, and also on an important quantity for determination of the SLD profile of an unknown non-magnetic thin film by polarized incident beam. For weakly absorbing substrates, the difference may be appreciable, especially in local minima of the reflectivity vs. neutron wave number, such that it must be considered to better interpretation of the measured data in neutron specular reflection experiments.